# Price Gouging or Market Forces?
# Fairness Perceptions of Price Hikes in the Pandemic

Avichai Snir
Department of Economics, Bar-Ilan University
Ramat-Gan 52900, ISRAEL
avichai.snir@gmail.com

Daniel Levy*
Department of Economics, Bar-Ilan University
Ramat-Gan 52900, ISRAEL,
Department of Economics, Emory University
Atlanta, GA 30322, USA,
ICEA, ISET at TSU, and RCEA
Daniel.Levy@biu.ac.il

Dudi Levy
Department of Economics, Bar-Ilan University
Ramat-Gan 52900, ISRAEL
dudilevi1985@gmail.com

Haipeng Allan Chen
Tippie College of Business, University of Iowa
Iowa City, IA 52242, USA
haipeng-chen@uiowa.edu

February 27, 2024

**Abstract**: We report the results of surveys we conducted in the US and Israel in 2020, a time when many prices increased following the spread of the COVID-19 pandemic. To assess respondents' perceptions of price increases, we focus on goods whose prices have increased during the pandemic, including some essential goods. Consistent with the principle of dual entitlement, we find that respondents perceive price increases as more acceptable if they are due to cost shocks than if they are due to demand shocks. However, we also find large differences across the two populations, as well as across goods.

**Disclosure Statement**: We have no potential conflict of interest.

* Corresponding author: Daniel Levy, Daniel.Levy@biu.ac.il

"For an economist, one of the most jarring sights during the [pandemic]…was the spectacle of bare shelves…There was no toilet paper or hand sanitizer…hospitals could not buy enough of the masks, gowns, and ventilators…What happened to the laws of supply and demand? Why didn't prices rise enough to clear the market…? A paper that I wrote with…Daniel Kahneman and Jack Knetsch…explored this problem. We found that the answer may be summed up with a single word, one you won't find in the standard supply-and-demand models: fairness."

Richard Thaler, "When the Law of Supply and Demand Isn't Fair,"
*New York Times*, May 24, 2020, p. 8

## 1. Introduction

When do people interpret a price increase as unfair? Blinder et al. (1998) and Anderson and Simester (2010) identify consumer-antagonization as one of the culprits.

By the principle of dual-entitlement (Kahneman et al. 1986a,b), consumers are entitled to their reference transaction-terms and firms to their reference profits. Consumers' perceptions of price increases also depend on whether the good's quality or production costs have changed, whether the price increase is uniform across consumers, etc.[1]

The pandemic offers an interesting setting for revisiting this question. During 2020, as COVID-19 was spreading, many prices increased significantly (Cabral and Xu 2021). There were even reports of price-gouging.[2]

We report the results of a survey we conducted in the U.S. and Israel, in 2020. In the survey, we asked respondents about their perception of price increases, focusing on goods whose prices have increased during the pandemic, including some essential goods.

We report five findings. First, Israeli respondents find price increases less acceptable than U.S. respondents. Second, consistent with the principle of dual-entitlement, for all products and in both countries, participants perceive supply-driven price increases as more acceptable than demand-driven price increases or price increases for unknown reasons. Third, in some cases, most respondents perceive the price increases as unacceptable regardless of their cause. Fourth, in the U.S., there is large variability across goods in the respondents' perception of the price increase fairness. Fifth, the respondents' attitudes towards price increases are related to the importance they attach to the goods, but more in the US than in Israel.

In section 2, we describe the methodology and the data. In section 3, we present the results. We conclude in section 4.

---

[1] See Urbany et al. (1989), Frey and Pommerehne (1993), Bolton et al. (2003), Xia et al. (2004), Kalapurakal et al. (1991), Leibbrandt (2020), Friedman and Toubia (2020), and Allender et al. (2021).
[2] For example, "Amazon has misled the public…about price increases during the pandemic. Numerous examples of price increases were found…some as much as 1,000%" (p. 4). This led to class-action lawsuits against Amazon." Source: https://www.hbsslaw.com/sites/default/files/case-downloads/amazon-price-gouging/2021-10-22-first-amended-complaint.pdf.

**2. Methodology and data**

In the U.S., we conducted a survey via Amazon mTurk among US residents.[3] In Israel, we conducted the survey via a business-news website, www.bizportal.co.il, and through social media forums.[4] We conducted the survey in November–December 2020, a period of significant restrictions. We have 904 respondents in the US and 1,043 in Israel. Table 1 presents summary statistics.

The survey contained 5 questions about goods whose prices have increased. The participants were asked to indicate for each whether it is *completely fair*, *acceptable*, *unfair*, or *very unfair* (Kahneman et al., 1986a). In the US, the goods chosen were facemasks, hand-sanitizers, toilet paper, chicken, and Dijon-mustard. The first three were chosen because they were essential during the pandemic, and the prices of all three had increased significantly. The price of chicken had also increased, but unlike the other products, no shortage of chicken was reported.[5] Dijon-mustard served as a control, because it experienced no supply- or demand-shocks. In Israel, we replaced toilet papers with eggs, because in Israel there was no shortage of toilet paper, but there were shortages of eggs.[6]

We employed three scenarios for each good. In one scenario, the price increase was due to an increase in demand. In the second, the reason for the price increase was supply shortages due to COVID-19 lockdowns. In the third, no reason was given. In all cases, the price increase was about 30%. Each respondent saw only one (randomly assigned) scenario for each question.

**3. Results**

Following Kahneman et al. (1986a), for each scenario we group the participants into two categories: those that judge the price increase as *acceptable* ("fair"/"acceptable") and those that judge the price increase as *unacceptable* ("unfair"/"very unfair"). Figure 1 shows the % of participants who thought the price-hikes were *unacceptable*.

We report five findings. First, Israelis find price increases less acceptable than

---

[3] The Appendix contains the survey form/text.
[4] No Helsinki committee permission was needed. The dataset is available upon request.
[5] See: https://time.com/5830178/meat-shortages-coronavirus/.
[6] See: www.haaretz.com/israel-news/.premium-amid-nationwide-shortage-israelis-scramble-for-eggs-1.8742526.

Americans. Across all scenarios and goods, 77.44% of the Israelis find price increases unacceptable, compared to 45.51% of the Americans. The difference is statistically significant (Wilcoxon-rank-sum $z = 36.71, p < 0.00$).

Second, consistent with the principle of dual entitlement, for all products and in both countries, participants perceive supply-driven price increases as more acceptable than demand-driven price increases or price increases for unknown reasons. In the U.S., across all goods, 45.71% of the respondents perceive a supply-driven price increase as unacceptable, compared to 56.27% that perceive a demand-driven price increase as unacceptable (Wilcoxon-rank-sum $z = 3.30, p < 0.00$), and 56.90% that perceive a price increase for unknown reasons as unacceptable (Wilcoxon-rank-sum $z = 3.49, p < 0.00$). The fairness perceptions of price increases that are demand-driven or that are due to unknown reasons are not significantly different from each other (Wilcoxon-rank-sum $z = 0.35, p > 0.72$).

Similarly, in Israel, across all goods, 67.52% of participants perceive a supply-driven price increase as unacceptable, compared to 82.41% that perceive a demand-driven price increase as unacceptable (Wilcoxon-rank-sum $z = 9.34, p < 0.00$), and 82.32% that perceive a price increase that is due to unknown reasons as unacceptable (Wilcoxon-rank-sum $z = 9.24, p < 0.00$). The fairness perceptions of price increases that are demand-driven or that are due to unknown reasons are not significantly different from each other (Wilcoxon-rank-sum $z = 0.07, p < 0.94$).

Third, in some cases, most of the respondents perceive the increases as unacceptable regardless of their cause. In the U.S., 55.30% and 67.67% of the respondents perceive supply-driven price increases of hand-sanitizers and toilet-paper, respectively, as unacceptable. Both values are significantly greater than 50% (Pearson-$\chi^2 = 3.19$ and 36.75, with $p < 0.08$ and $p < 0.00$, respectively). In Israel, for all goods, the perc-entage of the participants that perceive supply-driven price increases as unacceptable is always significantly greater than 50% (Pearson-$\chi^2 \geq 7.62$ in all cases, $p < 0.00$).

Fourth, in the U.S., there is large variability across goods in the respondents' perception of the price increase fairness. Combined across all 3 scenarios (demand driven, supply driven, and unknown reasons), 51.88%, 68.25%, 68.47%, 36.62%, and 39.60% of the participants view the price increases of facemasks, hand-sanitizers, toilet-papers, chicken and Dijon-mustard, respectively, as unacceptable. In Israel, the variability is smaller, but still nontrivial. Across the three scenarios, 70.71%, 78.66%, 82.65%, 75.74%, and 78.24% of the participants view the price increases of facemasks, hand-sanitizers, eggs, chicken, and Dijon-mustard, respectively, as unacceptable.

Fifth, to check whether the respondents' attitudes towards price increases are related to the importance they attach to the goods (Kalapurakal et al. 1991), we asked the respondents (*after* they have finished answering the fairness questions) to rate on a scale 1–5 the importance they attach to each good.

Table 2 summarizes the results. In both countries, we find large variation in the importance the respondents attach to the goods. In the U.S., toilet-papers are ranked as the most important with an average score of 4.48, while in Israel—the facemasks, with an average score of 3.39. In both countries, Dijon-mustard is ranked as the least important, with an average score of 1.98 in the U.S., and 1.60 in Israel. Thus, in both countries, the highest average score of importance is more than twice the lowest average score.

To examine the correlation between the fairness perceptions and the importance of the goods, we estimate random effects' linear regressions. To control for differences between respondents, we cluster the standard errors at the respondents' level. The dependent variable is a dummy that equals 1 if a price increase is unacceptable. The estimation results are reported in Table 3. In column 1, the independent variables are the importance score, dummies for goods, dummies for the reason for the price increase, a dummy for the Israeli sample, and an interaction of the dummy for the Israeli sample with the importance score.

We find that the coefficient of the importance score is positive $(\beta = 0.03, p < 0.00)$, suggesting that US respondents consider the importance of the goods when they assess the fairness of the goods' price increase: the more important the good, the more likely they are to perceive a price increase as unacceptable. The coefficient of the interaction of the importance score with the Israeli sample is negative $(\beta = -0.02, p < 0.05)$. The sum of the main effect and the interaction term is only marginally different from zero $(\chi^2 = 3.39, p < 0.07)$, suggesting that when assessing fairness of price increases, the goods' importance matters less for the Israeli respondents than for the U.S. respondents.

In column 2, we include additional controls: the respondents' age, a dummy for married respondents, a dummy for employed respondents, a dummy for respondents with academic degree, and a dummy for respondents that have taken at least one college-level economics course. Adding these variables does not change the coefficient or the significance of the importance score.

**4. Conclusion**

We conducted surveys in the U.S. and Israel, to assess people's attitudes towards price increases during the pandemic. Consistent with the principle of dual-entitlement, we find that respondents perceive supply-driven price increases as more acceptable than demand-driven price increases. In the U.S. respondents that attach greater importance to a good are more likely to respond that its price increase is unacceptable.

In Israel, most respondents perceive price increases during the pandemic as unacceptable regardless of the reason, perhaps because of the deep-rooted sentiments among Israelis that their country is expensive, and any price increase is unacceptable.[7] Indeed, prices in Israel are on average 20% higher than in the OECD countries (Avishay-Rizi and Ater 2021, 2022, and Hendel, Lach, and Spiegel 2017).[8]

Thus, although the principle of dual-entitlement applies in both the US and Israel, in times such as the pandemic shoppers may perceive price hikes as unfair, regardless of the reason for the price hike. Also, there are differences between the attitudes of consumers in different markets (Bolton et al. 2010). Israeli consumers perceive price-hikes as unacceptable regardless of how important a good is, whereas U.S. consumers' fairness perceptions of price-hikes depend more on the importance of the good.

---

[7] *Economist* (2021) ranked Tal-Aviv as one of the most expensive cities.
[8] Source: https://fs.knesset.gov.il/globaldocs/MMM/b42f5020-4ceb-e911-810f-00155d0af32a/2_b42f5020-4ceb-e911-810f-00155d0af32a_11_13731.pdf.

Table 1. Descriptive statistics of the US and Israeli participant samples

|  | US | Israel |
|---|---|---|
| % Women | 64.38% | 65.54% |
| % Married | 42.48% | 45.25% |
| % Employed | 77.54% | 74.40% |
| % Academics | 46.90% | 36.82% |
| % Studied economics | 59.49% | 46.60% |
| Average age | 38.60 | 32.37 |

Notes: % of women, married, employed (at least part time), academics (BA degree or higher), studied economics (at least one college-level course in economics), are their shares in the corresponding sample. The average age is the participants' average age.

Table 2. Importance scores of the goods

|  | US | Israel |
|---|---|---|
| Facemasks | 3.61 (1.311) | 3.39 (0.833) |
| Hand-sanitizers | 3.75 (1.197) | 2.96 (0.912) |
| Toilet-paper | 4.48 (0.853) |  |
| Eggs |  | 3.37 (0.732) |
| Chicken | 3.46 (1.174) | 3.08 (0.852) |
| Dijon-mustard | 1.98 (1.093) | 1.60 (0.736) |

Notes: The average responses to the question: "On a scale from 1–5, how important are _______ to you?" Standard deviations are given in parentheses.

Table 3. Perception of price increase fairness and the goods' importance

|  | (1) | (2) |
|---|---|---|
| Importance score | 0.03*** | 0.03*** |
|  | (0.006) | (0.006) |
| Importance score × Israeli sample | -0.02** | -0.02*** |
|  | (0.008) | (0.008) |
| Facemasks | -0.00 | -0.00 |
|  | (0.016) | (0.016) |
| Hand-sanitizers | 0.12*** | 0.12*** |
|  | (0.015) | (0.015) |
| Toilet-paper | 0.18*** | 0.18*** |
|  | (0.020) | (0.020) |
| Eggs | 0.06*** | 0.06*** |
|  | (0.017) | (0.017) |
| Chicken | -0.04*** | -0.04*** |
|  | (0.014) | (0.014) |
| Unknown reason for price increase | 0.13*** | 0.13*** |
|  | (0.010) | (0.010) |
| Price increase due to demand | 0.12*** | 0.12*** |
|  | (0.009) | (0.009) |
| Israeli sample | 0.33 | 0.34*** |
|  | (0.028) | (0.028) |
| Age |  | 0.00*** |
|  |  | (0.000) |
| Married |  | -0.02 |
|  |  | (0.016) |
| Employed |  | -0.02 |
|  |  | (0.016) |
| Academic |  | -0.03* |
|  |  | (0.015) |
| Taken economic course |  | -0.03* |
|  |  | (0.015) |
| Constant | 0.30*** | 0.29*** |
|  | (0.021) | (0.033) |
| $\chi^2$ | 1,032.3*** | 1,082.9*** |
| N | 9,233 | 9,233 |

Notes: Results of random effects regressions with standard errors clustered at the respondents' level. The dependent variable in both columns is a dummy that equals 1 if the respondent assessed a price increase as unacceptable and 0 otherwise. Importance score is the respondent's response to the question: "On a scale from 1-5, how important are _______ to you?" Importance score × Israeli sample is an interaction term between the importance score and a dummy for respondents participating in the Israeli survey. facemasks, hand-sanitizers, toilet-paper, eggs, and chicken are all product dummies that receive 1 if the product is facemasks, hand-sanitizers, toilet-paper, eggs, and chicken, respectively, and 0 otherwise. An unknown reason for price increase is a dummy that equals 1 if the respondent was not given a reason for the price increase, and 0 otherwise. Price increase due to demand is a dummy that equals 1 if the respondent was told that the reason for the price increase is an increase in demand, and 0 otherwise. Israeli sample is a dummy that equals 1 if the respondent took part in the Israeli survey, and 0 otherwise. Age is the respondent's age. Married is a dummy that equals 1 if the respondent is married and 0 otherwise. Employed is a dummy that equals if the respondent is employed full or part time and 0 otherwise. Academic is a dummy that equals 1 if the respondent has BA or higher degree. Taken economic course is a dummy that equals 1 if the participant has taken at least one college-level economics course. * $p < 10\%$, ** $p < 5\%$, *** $p < 1\%$.

Figure 1. Percentage of participants that find price increases unacceptable

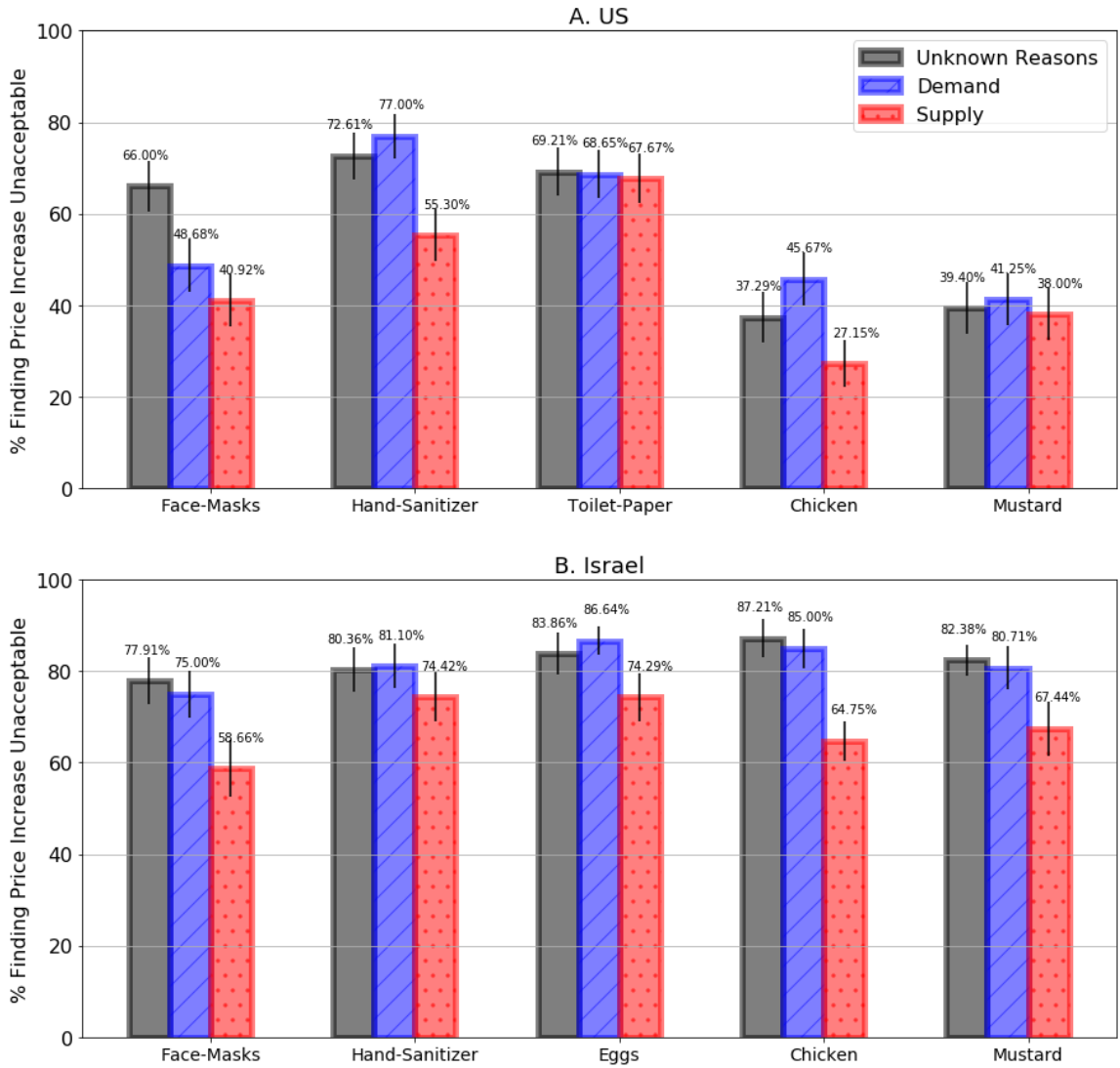

Notes: The small vertical lines indicate 2-standard error confidence bounds.

# Online Web Appendix

# Price Gouging or Market Forces?
# Fairness Perceptions of Price Hikes in the Pandemic

Avichai Snir
Department of Economics, Bar-Ilan University
Ramat-Gan 52900, ISRAEL
avichai.snir@gmail.com

Haipeng (Allan) Chen
Tippie College of Business, University of Iowa
Iowa City, IA 52242, USA
haipeng-chen@uiowa.edu

Daniel Levy
Department of Economics, Bar-Ilan University
Ramat-Gan 52900, ISRAEL,
Department of Economics, Emory University
Atlanta, GA 30322, USA,
ICEA, ISET at TSU, and RCEA
Daniel.Levy@biu.ac.il

Dudi Levy
Department of Economics, Bar-Ilan University
Ramat-Gan 52900, ISRAEL
dudilevi1985@gmail.com

## The Survey Questionnaire

Notes on the questionnaire

The section titles were not included in the questionnaire and thus they were not shown to the participants. We add them here (using a red-colored font) to make the structure of the questionnaire clearer. In the section of questions on the fairness of price increases, each participant saw only one scenario for each product. Thus, each participant saw one question on a price increase of face-masks, one question on a price increase of hand-sanitizers, one question on a price increase of toilet-papers, one question on a price increase of chicken, and one question on a price increase of Dijon mustard. The order of the questions, and the scenarios were assigned randomly, such that each participant saw at most two questions with the same scenario.

### Introduction of the Survey

Researchers at the University of Kentucky invite you to take part in a survey on consumer perceptions.

We invite you to assist us in learning about consumer perceptions of product prices. You may not get personal benefit from taking part in this research study, but your responses may help us understand more about purchasing decisions. Some people get satisfaction knowing they have contributed to research that may benefit others.

The survey will take about 5 minutes to complete. There are no known risks to participating in this study. Your response to the survey is confidential. This means no names will appear or be used on research documents. Names will not be used in presentations or publications either. The research team will not know that any information provided came from you, not even whether you participated in the study or not.

Your information collected for this study will NOT be used or shared for future research studies, not even if we remove the identifiable information like your age, gender, or race.

The software being used to collect your answers does collect your IP address. These are used to avoid duplicate answers, and the IP's are deleted from the data that the research team will use to make their analyses.

We hope to receive completed questionnaires from all participants. Your answers are important to us. You have a choice whether or not to complete the survey. You are also free to skip the questions if you do not want to answer them. If you do participate, you are free to stop at any time. If you decide to stop participating after you begin the survey, you can leave early and still get a completion code by contacting the research team at the email listed below.

Please be aware, we make every effort to keep your data safe when we get it from Qualtrics, the online survey company. Because of the nature of online surveys, as anything on the Internet, we cannot guarantee the confidentiality of the data while still on Qualtrics' servers. We cannot safeguard it while en route to either them or us either. It is also possible the raw data collected for research purposes will be used by Qualtrics. They may use it for marketing or reporting purposes. This depends on the company's Terms of Service and Privacy policies. If you have questions about the study, please feel free to ask. Our contact information is given below.

If you have complaints, suggestions, or questions about your rights, contact the staff in the University of Kentucky. Reach the Office of Research Integrity at 859-257-9428 or toll-free at 1-866-400-9428. Thank you in advance for your assistance with this important project.

Sincerely,

Dr. Daniel Chavez and Dr. Allan Chen

Department of Marketing and Supply Chain

University of Kentucky

PHONE: 859-257-8936

E-MAIL: daniel.chavez@uky.edu

**Assessing the fairness of price increases**

*Face masks – Price increase due to an unknown reason:*

**A store had been selling a box of fifty face masks for $4.89. Following the outbreak of COVID-19, the store raises prices to $6.39. Please rate this action as:**

    Fair        Acceptable        Unfair        Very unfair

*Face masks – Price increase due to a demand shock:*

**A store had been selling a box of fifty face masks for $4.89. Following the outbreak of COVID-19, the demand for face masks has greatly increased. The store raises prices to $6.39. Please rate this action as:**

    Fair        Acceptable        Unfair        Very unfair

*Face masks – Price increase due to a supply shock:*

**A store had been selling a box of fifty face masks for $4.89. Following the outbreak of COVID-19, several factories producing the masks were temporarily closed. The store raises prices to $6.39. Please rate this action as:**

    Fair        Acceptable        Unfair        Very unfair

*Hand sanitizer – Price increase due to an unknown reason:*

**A store had been selling a package of a dozen 8 oz hand sanitizers for $44.29. Following the outbreak of COVID-19, the store raises the price to $57.59. Please rate this action as:**

    Fair        Acceptable        Unfair        Very unfair

*Hand sanitizer – Price increase due to a demand shock:*

**A store had been selling a package of a dozen 8 oz hand sanitizers for $44.29. Following the outbreak of COVID-19, the demand for hand sanitizers has been greatly increased. The store raises the price to $57.59. Please rate this action as:**

    Fair        Acceptable        Unfair        Very unfair

*Hand sanitizer – Price increase due to a supply shock:*

**A store had been selling a package of a dozen 8 oz hand sanitizers for $44.29. Following the outbreak of COVID-19, several factories producing hand sanitizers were temporarily closed. The store raises the price to $57.59. Please rate this action as:**

    Fair        Acceptable        Unfair        Very unfair

*Toilet-Paper – Price increase due to an unknown reason:*

**A store had been selling a pack of 18 toilet paper rolls for $17.99. Following the outbreak of COVID-19, the store raises the price to $23.39. Please rate this action as:**

    Fair        Acceptable        Unfair        Very unfair

*Toilet-Paper – Price increase due to a demand shock:*

**A store had been selling a pack of 18 toilet paper rolls for $17.99. Following the outbreak of COVID-19, the demand for toilet paper has increased. The store raises the price to $23.39. Please rate this action as:**

    Fair        Acceptable        Unfair        Very unfair

*Toilet-Paper – Price increase due to a supply shock:*

**A store had been selling a pack of 18 toilet paper rolls for $17.99. Following the outbreak of COVID-19, several factories producing toilet paper were temporarily closed. The store raises the price to $23.39. Please rate this action as:**

    Fair      Acceptable      Unfair      Very unfair

*Chicken – Price increase due to an unknown reason:*

**A store was selling fresh whole chicken for $0.99 per pound. Following the outbreak of COVID-19, the store raises the price to $1.29 per pound. Please rate this action as:**

    Fair      Acceptable      Unfair      Very unfair

*Chicken – Price increase due to a demand shock:*

**A store was selling fresh whole chicken for $0.99 per pound. Following the outbreak of COVID-19, the demand for meat has increased. The store raises the price to $1.29 per pound. Please rate this action as:**

    Fair      Acceptable      Unfair      Very unfair

*Chicken – Price increase due to a supply shock:*

**A store was selling fresh whole chicken for $0.99 per pound. Following the outbreak of COVID-19, there has been temporary shutdown of several meat processing facilities. The store raises the price to $1.29 per pound. Please rate this action as:**

    Fair      Acceptable      Unfair      Very unfair

7*Dijon Mustard – Price increase due to an unknown reason:*

**A store had been selling 8 oz Original Dijon Mustard for $2.29. Following the outbreak of COVID-19, the store raises the prices to $2.99. Please rate this action as:**

    Fair       Acceptable       Unfair       Very unfair

*Dijon Mustard – Price increase due to a demand shock:*

**A store had been selling 8 oz Original Dijon Mustard for $2.29. Following the outbreak of COVID-19, the demand for mustard has increased. The store raises the prices to $2.99. Please rate this action as:**

    Fair       Acceptable       Unfair       Very unfair

*Dijon Mustard – Price increase due to a supply shock:*

**A store had been selling 8 oz Original Dijon Mustard for $2.29. Following the outbreak of COVID-19, several producers were temporarily closed. The store raises the prices to $2.99. Please rate this action as:**

    Fair       Acceptable       Unfair       Very unfair

**<u>Demographics</u>**

**Age**:

    _____

**Gender**:

    Male    Female

**Marital status**:

    Single    Married    Divorced    Widowed    Other

**Racial or ethnic identification**:

    Caucasian (other than Hispanic)

    Black or African American

    American Indian or Alaska Native

    Asian or Pacific Islander

    Hispanic

    Other

**What is the highest level of school you have completed or the highest degree you have received?**

    Less than high school degree

    High school graduate (high school diploma or equivalent including GED)

    Some college but no degree

    Associate degree in college (2-year)

    Bachelor's degree in college (4-year)

    Master's degree

    Doctoral degree

    Professional degree (JD, MD)

**Did either of your parents graduate from college?**

    No

    Yes, both parents

    Yes, mother only

    Yes, father only

**Do you work?**

    No

    Yes, part-time

    Yes, full time

**Have you taken any courses in economics?**

    No

    Yes, 1–2

    Yes, 3–4

    Yes, more than 4

**Do you describe yourself as**:

    Democrat

    Republican

    Independent

    Other / I do not know

**How often do you hang out with your friends (hours per week)?**

    1    2    3    4    5    6 or more

**Do you recycle any of the following**: **plastic, paper, newspaper, glass, batteries, aluminum?**

    Yes

    No

    Did you volunteer in any setting during the last 12 months?    Yes    No